\documentclass[letterpaper,english]{emulateapj}
\usepackage[T1]{fontenc}
\usepackage[latin9]{inputenc}
\usepackage{amssymb}
\usepackage{graphicx}

\makeatletter

\providecommand{\tabularnewline}{\\}

\newcommand{\source}{4U~1820$-$30}
\shorttitle{Photospheric Expansion from 4U 1820-30}
\shortauthors{Keek, et al.}

\makeatother

\usepackage{babel}
\begin{document}

\title{\emph{NICER} Detection of Strong Photospheric Expansion during a
Thermonuclear X-Ray Burst from 4U~1820$-$30}

\author{L.~Keek,\altaffilmark{1} Z.~Arzoumanian,\altaffilmark{2} D.~Chakrabarty,\altaffilmark{3}
J.~Chenevez,\altaffilmark{4} K.\,C.~Gendreau,\altaffilmark{2}
S.~Guillot,\altaffilmark{5,6} T.~Güver,\altaffilmark{7,8} J.~Homan,\altaffilmark{9,10}
G.\,K.~Jaisawal,\altaffilmark{4} B.~LaMarr,\altaffilmark{3} F.\,K.~Lamb,\altaffilmark{11,12}
S.~Mahmoodifar,\altaffilmark{2} C.\,B.~Markwardt,\altaffilmark{2}
T.~Okajima,\altaffilmark{2} T.\,E.~Strohmayer,\altaffilmark{2}
J.\,J.\,M.~in 't Zand\altaffilmark{10}}

\altaffiltext{1}{Department of Astronomy, University of Maryland, College Park, MD 20742, USA}
\altaffiltext{2}{X-ray Astrophysics Laboratory, Astrophysics Science Division, NASA/GSFC, Greenbelt, MD 20771, USA}
\altaffiltext{3}{MIT Kavli Institute for Astrophysics and Space Research, Massachusetts Institute of Technology, Cambridge, MA 02139, USA}
\altaffiltext{4}{National Space Institute, Technical University of Denmark, Elektrovej 327-328, DK-2800 Lyngby, Denmark}
\altaffiltext{5}{CNRS, IRAP, 9 avenue du Colonel Roche, BP 44346, F-31028 Toulouse Cedex 4, France}
\altaffiltext{6}{Université de Toulouse, CNES, UPS-OMP, F-31028 Toulouse, France}
\altaffiltext{7}{Department of Astronomy and Space Sciences, Science Faculty, Istanbul University, Beyaz\i t, 34119 Istanbul, Turkey}
\altaffiltext{8}{Istanbul University Observatory Research and Application Center, Beyaz\i t, 34119 Istanbul, Turkey}
\altaffiltext{9}{Eureka Scientific, Inc., 2452 Delmer Street, Oakland, CA 94602, USA}
\altaffiltext{10}{SRON, Netherlands Institute for Space Research, Sorbonnelaan 2, 3584 CA Utrecht, The Netherlands}
\altaffiltext{11}{Center for Theoretical Astrophysics and Department of Physics, University of Illinois at Urbana-Champaign, 1110 West Green Street, Urbana, IL 61801-3080, USA}
\altaffiltext{12}{Department of Astronomy, University of Illinois at Urbana-Champaign, 1002 West Green Street, Urbana, IL 61801-3074, USA}

\email{lkeek@umd.edu}
\begin{abstract}
 The \emph{Neutron Star Interior Composition Explorer} (\emph{NICER})
on the International Space Station observed strong photospheric expansion
of the neutron star in \source{} during a Type I X-ray burst. A thermonuclear
helium flash in the star's envelope powered a burst that reached the
Eddington limit. Radiation pressure pushed the photosphere out to
$\sim200\,\mathrm{km}$, while the blackbody temperature dropped to
$0.45\,\mathrm{keV}$. Previous observations of similar bursts were
performed with instruments that are sensitive only above $3\,\mathrm{keV}$,
and the burst signal was weak at low temperatures. \emph{NICER}'s
$0.2-12\,\mathrm{keV}$ passband enables the first complete detailed
observation of strong expansion bursts. The strong expansion lasted
only $0.6\,\mathrm{s}$, and was followed by moderate expansion with
a $20\,\mathrm{km}$ apparent radius, before the photosphere finally
settled back down at $3\,\mathrm{s}$ after the burst onset. In addition
to thermal emission from the neutron star, the \emph{NICER} spectra
reveal a second component that is well fit by optically thick Comptonization.
During the strong expansion, this component is six times brighter
than prior to the burst, and it accounts for $71\%$ of the flux.
In the moderate expansion phase, the Comptonization flux drops, while
the thermal component brightens, and the total flux remains constant
at the Eddington limit. We speculate that the thermal emission is
reprocessed in the accretion environment to form the Comptonization
component, and that changes in the covering fraction of the star explain
the evolution of the relative contributions to the total flux.
\end{abstract}


\section{Introduction}

Located in the globular cluster NGC~6624, \source{} is an ultracompact
X-ray binary with an orbital period of $11.4\,\mathrm{minutes}$ \citep{1987Stella}.
Material of predominantly helium composition is accreted onto the
neutron star \citep[e.g.,][]{2003Cumming}, where runaway thermonuclear
burning powers the X-ray bursts that have been observed from this
source since 1975 (\citealt{Grindlay1976}; for a recent review, see
\citealt{Galloway2017Review}). In the absence of hydrogen, nuclear
burning proceeds rapidly, unhindered by waiting points from weak decays
\citep[e.g.,][]{Weinberg2006}. Most fuel, therefore, burns at the
onset, and a high peak luminosity is reached quickly. When the luminosity
exceeds the Eddington limit, radiation pressure exceeds the gravitational
pull, and photospheric radius expansion (PRE) is the result \citep{Grindlay1980pre}.
During PRE, the luminosity remains near the Eddington limit. An increase
in the emitting area due to expansion is, therefore, accompanied by
a decrease of the photospheric temperature.

For most PRE bursts the observed blackbody radius increases by a factor
of a few, whereas a small subset of PRE bursts exhibit strong expansion
in excess of a factor $10$ \citep[e.g.,][]{Galloway2008catalog}.
A radius increase of a factor $>100$ is referred to as ``superexpansion''
\citep{Zand2010}. Strong expansion may drive a wind from the neutron
star \citep[e.g.,][]{Ebisuzaki1983,Paczynski1986}, and provide opportunities
to constrain the neutron star's compactness \citep[e.g.,][]{Paradijs1987,Zand2010}.

With instruments like the Proportional Counter Array \citep[PCA;][]{Jahoda2006}
on the \emph{Rossi X-ray Timing Explorer} (RXTE) the temperature decrease
during the expansion leads to a substantial loss of signal out of
the $3-60\,\mathrm{keV}$ passband, producing a characteristic dip
in the burst light curve \citep{Paczynski1983pre}. For the strongest
expansion, the spectral parameters could not be reliably measured.
The passband of the \emph{Neutron Star Interior Composition Explorer}
(\emph{\facility{NICER}}; \citealp{Gendreau2017}) on the International
Space Station, however, extends down to $0.2\,\mathrm{keV}$, making
it an ideal instrument to study strong expansion at high time resolution.
Older instrumentation with a similar passband has observed bursts
with at most moderate expansion \citep[e.g.,][]{Zand2013}.

In this Letter we study the first strong PRE burst from \source{}
detected with \emph{NICER}. The instrument and observations are introduced
in Section~\ref{sec:Observations-and-Spectral}. Time-resolved spectroscopy
shows that \emph{NICER} can track the thermal burst emission as well
as a Comptonization component throughout the expansion phase (Section~\ref{sec:Results}).
For the first time we can test models of expansion and wind generation
(Section~\ref{sec:Discussion}), and we discuss the prospects for
future \emph{NICER} studies of strong radius expansion bursts (Section~\ref{sec:Conclusions-and-Outlook}).

\section{Observations}

\label{sec:Observations-and-Spectral} 

In August 2017, \source{} was in the hard spectral state (Section~\ref{sec:pers_spectra}),
and over $2.2$~days \emph{NICER} observed the source for a total
good exposure of $60.9$~ks. Five bursts were detected, each with
a short duration ($\sim5\,\mathrm{s}$) and a high peak count rate
($\sim2\times10^{4}\,\mathrm{c\,s^{-1}}$). We perform a detailed
analysis of the first burst in ObsID~1050300108 on MJD~57994.37115
(2017 August 29). 

The X-ray Timing Instrument \citep[XTI;][]{Gendreau2016} on board
\emph{NICER} employs $56$ co-aligned X-ray concentrator optics and
silicon-drift detectors \citep{Prigozhin2012}, with $52$ in operation.
This configuration enables the detection of X-ray photons in the $0.2-12\,\mathrm{keV}$
passband at high time resolution and $<100\,\mathrm{eV}$ energy resolution,
with a peak effective area of $1900\,\mathrm{cm^{2}}$ at $1.5\,\mathrm{keV}$.
We create XTI spectra with \textsc{Heasoft} version 6.22.1 and \textsc{Nicerdas}
2017-09-06\_V002. The spectra are analyzed with \textsc{Xspec} 12.9.1p
\citep{Arnaud1996} and version 0.06 of the \emph{NICER} response
files. As background we use the blank-field spectrum of \citet{Keek2018},
which is appropriate for the conditions of our observation with a
low particle background and the ISS being on the night-side of the
Earth. For our analysis, we group neighboring spectral bins to ensure
a minimum of $15$ counts per bin.

\section{Results}

\label{sec:Results}

\subsection{Persistent Emission}

\label{sec:pers_spectra}

The burst happened near the start of the \emph{NICER} pointing. Therefore,
we characterize the persistent emission using a $2145\,\mathrm{s}$
interval at the end of the $2294\,\mathrm{s}$ pointing. We extract
a spectrum in the $0.3-9.0\,\mathrm{keV}$ band with $3.9\times10^{6}$~counts
(see Figure~\ref{fig:spectra} below). Following the broadband analysis
by \citet{Costantini2012}, we fit the persistent emission with a
combination of a Planck model (\texttt{bbodyrad} in \textsc{Xspec})
and a Comptonization component (\texttt{compTT}; \citealt{1994Titarchuk}).
The Tübingen-Boulder model (\texttt{TBabs}) for interstellar absorption
is employed with abundances from \citet{Wilms2000}. The fit exhibits
residuals near the instrumental edges around $0.5\,\mathrm{keV}$
and $2.3\,\mathrm{keV}$. This indicates a small shift in the gain
that is not included in the current model of the instrument response.
We use \textsc{Xspec}'s gain model to optimize the energy scale of
our data, finding a gain offset of $4.5\,\mathrm{eV}$ and slope of
$1.008$, which is small with respect to the $\sim100\,\mathrm{eV}$
energy resolution. This substantially improves the fit, but some features
remain in the residuals. Further improvements to the response model
are needed to fully resolve these issues. In this study we regard
it a systematic uncertainty. A goodness of fit per degree of freedom
of $\chi_{\nu}^{2}=1.0$ ($\nu=860$) is obtained by adding a $1.5\%$
error in quadrature to the statistical error of each data point. 

\begin{table}

\caption{\label{tab:persistent}Best fit to the persistent spectrum}

\begin{centering}
\begin{tabular}{cc}
\hline 
Parameter & Value\tabularnewline
\hline 
\multicolumn{2}{c}{\texttt{TBabs}}\tabularnewline
$N_{\mathrm{H}}\,(\mathrm{10^{21}\,cm^{-2}})$ & $2.53\pm0.02$\tabularnewline
\hline 
\multicolumn{2}{c}{\texttt{bbodyrad}}\tabularnewline
$kT\,(\mathrm{keV})$ & $0.561\pm0.007$\tabularnewline
$R_{\mathrm{bb}}\,(\mathrm{km\,at\,8.4\,kpc})$ & $21.2\pm0.5$\tabularnewline
\hline 
\multicolumn{2}{c}{\texttt{compTT}}\tabularnewline
$kT_{0}\,(\mathrm{keV})$ & $0.04\pm0.02$\tabularnewline
$kT\,(\mathrm{keV})$ & $2.69\pm0.06$\tabularnewline
$\tau$ & $7.60\pm0.12$\tabularnewline
$K_{\mathrm{compTT}}\,(\mathrm{c\,s^{-1}\,keV^{-1}})$ & $2.4\pm0.9$\tabularnewline
\hline 
$F_{0.3-9\,\mathrm{keV}}(10^{-9}\,\mathrm{erg\,s^{-1}\,cm^{-2}})$ & $6.89\pm0.02$\tabularnewline
\end{tabular}
\par\end{centering}
\end{table}
 The best-fitting parameter values and $1\sigma$ uncertainties are
presented in Table~\ref{tab:persistent}. The absorption column,
$N_{\mathrm{H}}$, is consistent within $1\sigma$ with the mean value
obtained from \emph{Chandra }grating spectra \citep{Guver2010}. We
use the \texttt{cflux} model to determine the unabsorbed Comptonization
flux in the $0.3-9.0\,\mathrm{keV}$ band to be $(6.31\pm0.03)\times10^{-9}\,\mathrm{erg\,s^{-1}\,cm^{-2}}$,
and by extrapolation to the $0.001-100\,\mathrm{keV}$ band we derive
an unabsorbed bolometric flux of $(8.86\pm0.09)\times10^{-9}\,\mathrm{erg\,s^{-1}\,cm^{-2}}$.

From the normalization of the blackbody we derive the apparent radius,
$R_{\mathrm{bb}}=21.2\pm0.5\,\mathrm{km}$, under assumption of an
isotropically emitting spherical surface at a distance of $8.4\,\mathrm{kpc}$
\citep{Valenti2004}. This is larger than expected for the neutron
star, and may represent emission from the inner disk. We caution that
the best-fit blackbody parameters may be sensitive to the energy $kT_{0}$
of the seed photons being Comptonized. Including the blackbody component,
the total unabsorbed persistent flux is $(6.89\pm0.02)\times10^{-9}\,\mathrm{erg\,s^{-1}\,cm^{-2}}$
($0.3-9.0\,\mathrm{keV}$) or $(9.54\pm0.09)\times10^{-9}\,\mathrm{erg\,s^{-1}\,cm^{-2}}$
($0.001-100\,\mathrm{keV}$). 

As a measure of the Eddington flux, we find from the Multi-Instrument
Burst Archive \citep[MINBAR; e.g.,][]{Galloway2008catalog} the weighted
mean of the bolometric peak blackbody flux of $67$ PRE bursts observed
with \emph{RXTE}/PCA from \source{}: $F_{\mathrm{Edd}}=(6.08\pm0.06)\times10^{-8}\,\mathrm{erg\,s^{-1}\,cm^{-2}}$.
Therefore, the total persistent flux measured with \emph{NICER} is
$(0.157\pm0.002)\,F_{\mathrm{Edd}}$. 

\subsection{Burst Light Curve}

\label{subsec:Burst-Light-Curve}

\begin{figure}
\includegraphics{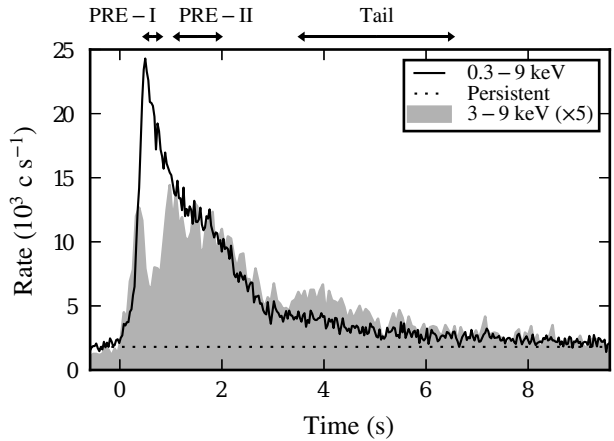}

\caption{\emph{\label{fig:lcv}}Photon count rate as a function of time in
the $0.3-9\,\mathrm{keV}$ passband at $30\,\mathrm{ms}$ resolution.
The top of the shaded region marks the $3-9\,\mathrm{keV}$ count
rate for comparison (scaled by a factor of $5$). The dotted line
indicates the persistent count rate measured at the end of the observation.
On top, three time intervals are indicated for spectroscopy (Figure~\ref{fig:spectra}).}
\end{figure}
 The burst light curve reaches a peak count rate of $2.4\times10^{4}\,\mathrm{c\,s^{-1}}$
in the $0.3-9\,\mathrm{keV}$ band (Figure~\ref{fig:lcv}). When
we consider only the photons with energies in excess of $3\,\mathrm{keV}$,
where past instruments such as\emph{ RXTE}/PCA were sensitive, a dip
appears in the light curve. Such a dip has been found to be the characteristic
signature of PRE \citep{Grindlay1980pre}. \emph{NICER}'s coverage
of the soft X-ray band provides a complementary view of this bright
burst phase, and the count rate spikes. The XTI's modularity accommodates
these large count rates without pile-up or telemetry issues. The high
count rate is maintained for a short duration ($\sim0.5\,\mathrm{s}$),
and is quickly reduced to $\sim1.2\times10^{4}\,\mathrm{c\,s^{-1}}$.
The latter level appears as a $\sim0.5\,\mathrm{s}$ ``plateau'',
after which the count rate returns to the persistent level on a timescale
of $\sim8\,\mathrm{s}$. 

No burst oscillations have been detected during this burst.

\subsection{Burst Spectra}

\label{subsec:Burst}

\begin{figure}
\includegraphics{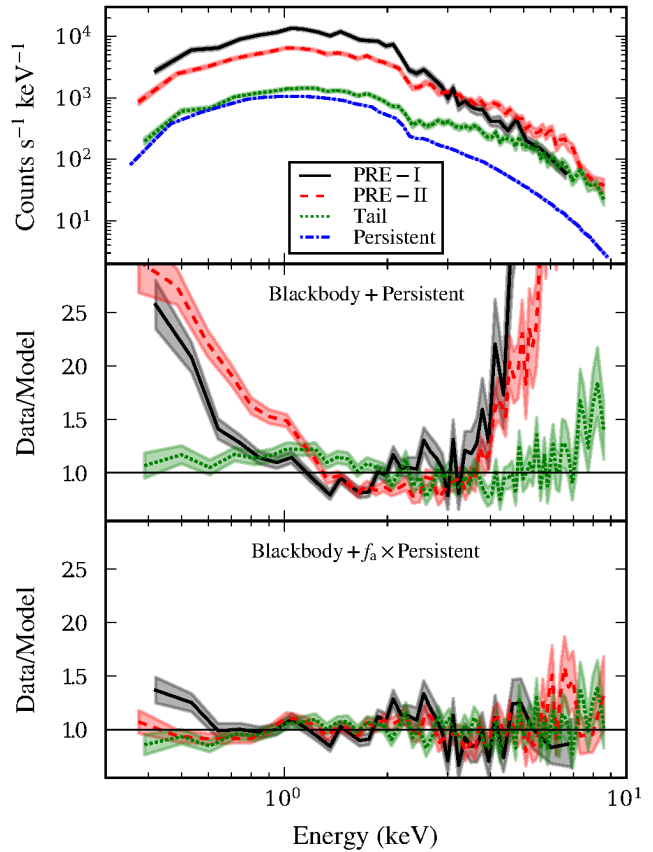}

\caption{\emph{\label{fig:spectra} }Top panel: X-ray spectra from three intervals
in the burst (Figure~\ref{fig:lcv}) as well as the persistent emission.
The spectra are rebinned to a resolution of at most $50\,\mathrm{eV}$,
and shaded bands indicate the $1\sigma$ uncertainty ranges. Middle
panel: ratio of the burst spectra and best-fitting models with an
absorbed blackbody and fixed persistent Comptonization component.
Bottom panel: ratios for similar fits where the persistent Comptonization
component is scaled with independent factors $f_{\mathrm{a}}$.}
\end{figure}
 To investigate the importance of the different spectral components
during the burst, we extract spectra from three time intervals. We
select an interval of $0.33\,\mathrm{s}$ around the peak in the count
rate (``PRE-I''; Figures~\ref{fig:lcv},~\ref{fig:spectra}),
$0.9\,\mathrm{s}$ in the subsequent plateau (``PRE-II''), and $3\,\mathrm{s}$
in the tail (``Tail''). We first fit the spectra with the usual
burst model, where we keep the Comptonization component parameters
fixed at the values from Table~\ref{tab:persistent} and employ the
blackbody component to model the thermal burst emission. The absorption
column, $N_{\mathrm{H}}$, is fixed to the best-fitting value from
the persistent emission. Furthermore, we use the same gain corrections
as in Section~\ref{sec:pers_spectra}. This spectral model does not
provide a satisfactory description of the data ($\chi_{\nu}^{2}(\nu)=2.38\,(236),\,4.08\,(343),\,1.21\,(358)$
for PRE-I, PRE-II, and the tail, respectively), with excesses appearing
at both ends of the passband (Figure~\ref{fig:spectra} middle panel).
Next, we allow for a scaling factor, $f_{\mathrm{a}}$, to change
the normalization of the Comptonization component \citep[e.g.,][]{Worpel2013}.
This scaling is a purely phenomenological assumption. Using this scaling,
the fits are vastly improved ($\chi_{\nu}^{2}(\nu)=1.37\,(235),\,1.12\,(342),\,0.94\,(357)$
for PRE-I, PRE-II, and the tail, respectively), and provide a reasonable
description of the spectra (Figure~\ref{fig:spectra} bottom panel).
We employ this ``$f_{\mathrm{a}}$ model'' in time resolved spectroscopy.
In all these fits, the seed and electron temperatures of the Comptonization
component during the burst are fixed at the values found for the persistent
emission (Table~\ref{tab:persistent}). Although this produces good
fits, it is not a self-consistent physical model.

Instead of Comptonization, we also tested the same model for disk
reflection that was successfully applied to the 1999 superburst from
\source{} \citep{Ballantyne2004}. It fails to provide an adequate
fit, especially for the excess at $\gtrsim3\,\mathrm{keV}$ (Figure~\ref{fig:spectra}
middle). We find a $90\%$ upper limit to the reflection fraction
of $f_{\mathrm{refl}}=0.22$, which is close to the value measured
in the superburst.

\subsection{Time-resolved Spectroscopy}

\label{subsec:Time-Resolved-Spectroscopy}

\begin{figure}
\includegraphics{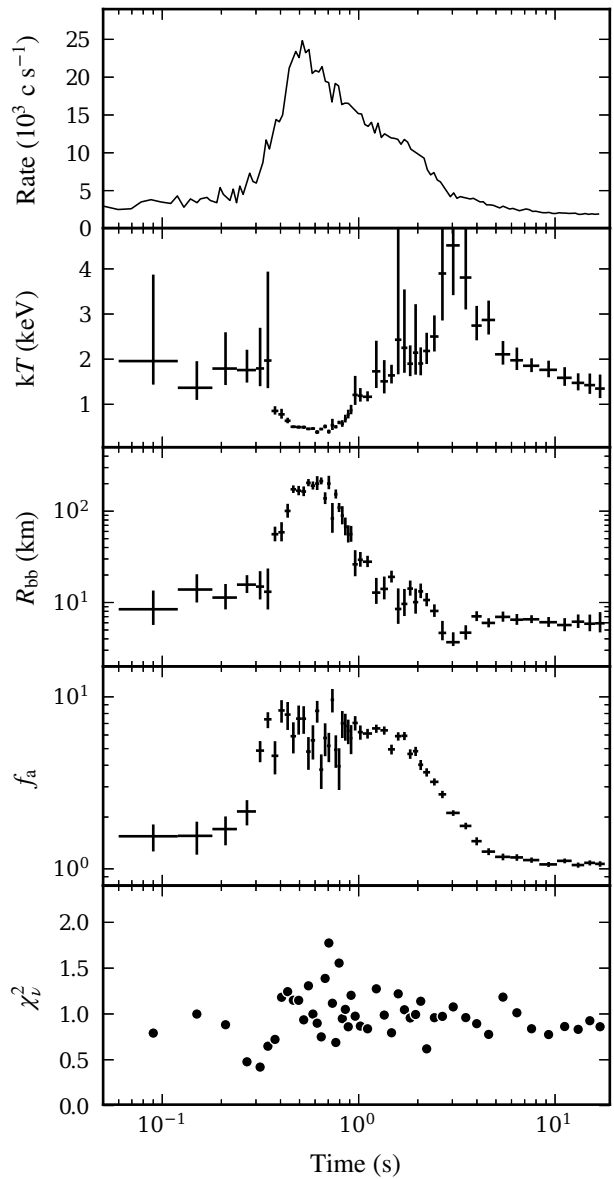}

\caption{\emph{\label{fig:fa_time_resolved} }Results of time-resolved spectroscopy.
The top panel shows the count rate, the bottom panel shows the goodness
of fit $\chi_{\nu}^{2}$, and the other panels show the best-fitting
values with $1\sigma$ errors for the blackbody temperature $kT$,
radius of the emitting area $R_{\mathrm{bb}}$ (assuming a distance
of $8.4\,\mathrm{kpc}$), and scaling factor for the persistent Comptonization
flux $f_{\mathrm{a}}$.}
\end{figure}

We extract spectra from $52$ time intervals with durations between
$0.03\,\mathrm{s}$ where the count rate peaks and $1.92\,\mathrm{s}$
in the tail. The spectra contain on average $1239$ counts each ($6.4\times10^{4}$
counts in total). 

The spectra are fit with the $f_{\mathrm{a}}$ model (Section~\ref{subsec:Burst}).
Strong photospheric expansion is apparent: $R_{\mathrm{bb}}$ reaches
a maximum of $R_{\mathrm{bb}}=190\pm10\,\mathrm{km}$, accompanied
by a minimum in the temperature of $kT=0.449\pm0.013\,\mathrm{keV}$
(weighted means of six bins around the extrema). The phase of strong
expansion lasts only $\sim0.6\,\mathrm{s}$. The subsequent decrease
in radius slows down, however, for $\sim1\,\mathrm{s}$ during a plateau
of moderate expansion with $R_{\mathrm{bb}}=13.5\pm1.3\,\mathrm{km}$
and $kT=1.7\pm0.2\,\mathrm{keV}$. The end of the PRE phase (``touchdown'')
is marked by a peak in $kT$ around $3\,\mathrm{s}$, coinciding with
a brief dip in $R_{\mathrm{bb}}$ \citep[see also, e.g.,][]{Zhang2013},
after which $R_{\mathrm{bb}}$ remains stable ($R_{\mathrm{bb}}=6.3\pm0.3\,\mathrm{km}$).
$R_{\mathrm{bb}}$ in the strong and moderate PRE phases are, respectively,
$30\pm2$ and $2.1\pm0.2$ times the value in the tail. At touchdown,
the temperature is near $kT\simeq3\,\mathrm{keV}$, but the uncertainties
are large. For this value of $kT$, the peak of the photon counts
spectrum is near $6\,\mathrm{keV}$, where the \emph{NICER} effective
area is substantially reduced with respect to its peak around $\sim1.5\,\mathrm{keV}$.

At the burst onset there is a strong rise in $f_{\mathrm{a}}$ to
a mean value of $f_{\mathrm{a}}=5.99\pm0.12$. During the moderate
expansion phase, $f_{\mathrm{a}}$ begins to decline, returning to
unity (the persistent level) in the burst tail.

\begin{figure}
\includegraphics{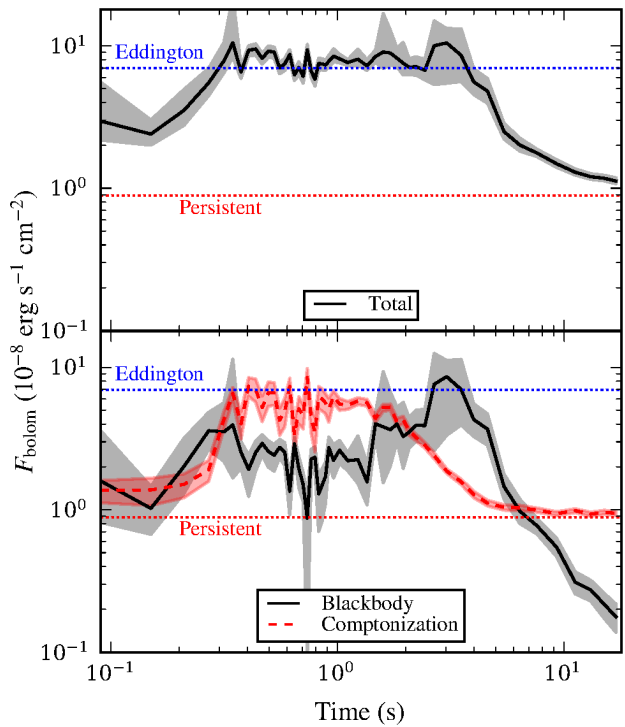}

\caption{\emph{\label{fig:flux} }Unabsorbed bolometric flux as a function
of time. Top panel: total flux (the shaded band indicates the $1\sigma$
uncertainty range), with horizontal dotted lines indicating the persistent
flux outside of the burst and the \emph{RXTE}/PCA Eddington flux (\emph{NICER}
persistent flux included). Bottom panel: flux for both components
separately.}
\end{figure}

The total unabsorbed bolometric flux peaks at a plateau with a weighted
mean value of $(7.52\pm0.12)\times10^{-8}\,\mathrm{erg\,s^{-1}\,cm^{-2}}$,
which is maintained up to touchdown (Figure~\ref{fig:flux} top).
Subtracting the persistent emission, it is $8.0\%$ larger than $F_{\mathrm{Edd}}$
for the \emph{RXTE}/PCA bursts (Section~\ref{sec:pers_spectra}).
This falls within the typical observed variations in the peak fluxes
of PRE bursts of the same source \citep[e.g.,][]{Kuulkers2003}. Integrating
the flux above the persistent level, we find a total burst fluence
of $(4.0\pm1.1)\times10^{-7}\,\mathrm{erg\,cm^{-2}}$. 

When we regard the two spectral components separately, the blackbody
fluence is $(2.8\pm1.0)\times10^{-7}\,\mathrm{erg\,cm^{-2}}$, and
the Comptonization fluence in excess of the persistent level is $(1.2\pm0.2)\times10^{-7}\,\mathrm{erg\,cm^{-2}}$.
Therefore, a fraction of $0.29\pm0.10$ of the total burst fluence
is in the Comptonization component. Prior to $t\simeq1.3\,\mathrm{s}$,
the fraction is $0.71\pm0.03$. Subsequently, the blackbody flux increases
up to the touchdown, whereas the Comptonization flux decreases. At
$t=3\,\mathrm{s}$ the flux fraction for Comptonization is $0.18\pm0.05$.

After the touchdown at $t=3\,\mathrm{s}$, the blackbody flux drops
as a power law. Following the prescription of \citet{Zand2017}, we
fit a power law to the blackbody flux decline,  and find an index
$\Gamma=2.0\pm0.2$, which is typical for short bursts from helium-rich
fuel \citep{Zand2014,Zand2017}. Similarly, we determine the power-law
slope of the decay of the Comptonization component in the $2-4\,\mathrm{s}$
time interval: $\Gamma=1.46\pm0.07$.

\section{Discussion}

\label{sec:Discussion}

\subsection{Comparison to Expansion and Wind Models}

\label{subsec:Comparison-to-Expansion}

The 1980s saw the first models of PRE and radiation-driven winds for
the most luminous bursts \citep[e.g.,][]{Ebisuzaki1983,Paczynski1983pre}.
\citet{Paczynski1986} created General Relativistic models that allowed
them to probe strong expansion, and they remarked ``\emph{Most likely,
the stars with winds are too cool to be detectable with the existing
X-ray instruments.}'' With the \emph{NICER} observations of \source{}
we finally have the soft X-ray coverage at high time resolution to
test these models and investigate the strong radius expansion regime.
Their model with mass outflow rate $\dot{M}_{0}=10^{17.5}\,\mathrm{g\,s^{-1}}$
predicts a photospheric radius that is within $1\sigma$ of the maximum
$R_{\mathrm{bb}}=190\pm10\,\mathrm{km}$ that we derive. The predicted
photospheric temperature of $0.48\,\mathrm{keV}$ is close to our
measured minimum of $kT=0.449\pm0.013\,\mathrm{keV}$. 

The model predicts an outflow velocity of $\simeq790\,\mathrm{km\,s^{-1}}$.
This value is strongly dependent on the outer boundary conditions,
such as the optical depth where the observed signal originates: different
conditions give velocities in the range of $700-5000\,\mathrm{km\,s^{-1}}$.
The large change in $R_{\mathrm{bb}}$ within $0.12\,\mathrm{s}$
at the burst onset suggests an expansion of $(1.5\pm0.4)\times10^{3}\,\mathrm{km\,s^{-1}}$,
which may be a lower limit on the outflow velocity.

In comparing our observations to these models, we included neither
the systematic uncertainties from the source distance, anisotropy
factors \citep[e.g.,][]{He2016}, and color corrections of the blackbody
parameters \citep[e.g.,][]{Suleimanov2012}, nor the gravitational
redshift. Nevertheless, it is encouraging that the observed properties
of the strong PRE phase are well described by these relatively simple
models. Predictions from models with improved boundary conditions
\citep{Joss1987} and radiation transport \citep{Nobili1994} differ
by factors of only a few.

\subsection{Comparison to\emph{ }Previous Bursts: Superexpansion}

During superexpansion bursts observed by \emph{RXTE}/PCA, the thermal
emission moved below the $3-60\,\mathrm{keV}$ passband, and even
the persistent component disappeared \citep{Zand2010}. This was observed
from \source{} at a low persistent flux. During bursts at $\sim2-3$
times higher persistent flux, the source remained detectable, even
though the bursts had a similar fluence and duration. \citet{Zand2012}
speculated that higher accretion rates affect photospheric expansion
through the ram pressure of the infalling material. The \emph{NICER}
burst occurred at a relatively high persistent flux, which corresponds
to a mass accretion rate of $\dot{M}=(2.96\pm0.04)\times10^{17}\,\mathrm{g\,s^{-1}}$
(for hydrogen-deficient material and a $10\,\mathrm{km}$ neutron
star radius). This is close to the model-predicted outflow rate $\dot{M}_{0}$.
The ram pressure from accretion may have been important in setting
the extent of the expansion, whereas superexpansion occurs only at
lower accretion rates and ram pressures \citep{Zand2012}. To accurately
predict the expansion behavior, future models need to include the
accretion flow, and compare its ram pressure to that of the outflow.

\subsection{Enhanced Comptonization Emission}

\label{subsec:Enhanced-Comptonized-Emission}

The Comptonization component in our phenomenological model becomes
six times brighter during the burst. We discuss potential interpretations
of this component and of its time evolution.

Model spectra for strongly expanded atmospheres deviate from a blackbody
due to Comptonization and free-free absorption and emission \citep[e.g.,][]{1994TitarchukPRE}.
Our observed Comptonization component is substantially brighter than
the model predictions, and may instead be produced by reprocessing
of burst emission in the accretion environment. For example, a spreading
layer of accreted material could cover a substantial fraction of the
stellar surface during PRE \citep[e.g.,][]{Kajava2017}, and its spectrum
is thought to be well described by optically thick Comptonization
emission with a temperature of $kT\simeq2.5\,\mathrm{keV}$ \citep{Suleimanov2006,Revnivtsev2013},
similar to our \texttt{compTT} component (Table~\ref{tab:persistent}).

Alternatively, the burst emission could undergo Compton scattering
in the disk or corona. In the strong PRE phase, the Comptonization
component contributes $71\%$ of the flux. For such a large fraction
of the neutron star's thermal emission to be intercepted, the disk
or its corona must have a large scale height close to the star \citep[e.g.,][]{He2016}.
The flux fraction remains constant in this phase, despite large variations
in $R_{\mathrm{bb}}$. The fraction only decreases in the moderate
PRE phase, once $R_{\mathrm{bb}}$ drops below $\sim20\,\mathrm{km}$.
If the optically thick disk/corona is truncated at this radius, the
whole neutron star is revealed when the star's atmospheric radius
becomes smaller than the disk's inner radius. A truncated disk is
expected in the hard spectral state \citep[see also the discussion in][]{Zand2012}.
In this scenario, part of the neutron star is covered during the strong
PRE phase, and our measurement of $R_{\mathrm{bb}}$ represents the
visible fraction of the neutron star surface. If the total flux were
produced by the blackbody component, the larger normalization suggests
a maximum expansion of $R\mathrm{_{bb}}=350\pm40\,\mathrm{km}$.

If the thermal emission is Comptonized, one expects the spectral shape
of the Comptonization component to change with the blackbody temperature.
Nevertheless, we obtain good fits with a fixed shape, despite changes
in the blackbody temperature and the Compton component luminosity.
This is similar to other studies, where the shape of the enhanced
component matches the persistent spectrum outside of the burst \citep[see, e.g., Figure 4 of][]{Zand2013}.
Further burst observations with \emph{NICER} and additional theory/spectral
modeling efforts will be instrumental in finding a more self-consistent
physical description of superexpansion bursts.

\section{Conclusions and Outlook}

\label{sec:Conclusions-and-Outlook}

We have presented the first strong PRE burst from \source{} observed
with \emph{NICER}. Because of \emph{NICER}'s soft-band coverage, the
properties of the thermal emission could be traced even when the blackbody
temperature dropped to $0.45\,\mathrm{keV}$ and the radius increased
by a factor $30$. Furthermore, in the soft band a Comptonization
component was detected that accounts for up to $71\%$ of the energy
flux. Because the total flux during PRE remained at the Eddington
limit, we speculate that part of the neutron star was covered by the
accretion environment, and the thermal emission from the neutron star
was Comptonized. At the end of the PRE phase, the neutron star was
uncovered due to geometrical changes. 

The properties of the blackbody match to first order the predictions
from models of steady-state outflows. Currently, no detailed time-dependent
models exist, nor models that include the interaction with the accretion
environment \citep[e.g.,][]{Ballantyne2005}. Our findings indicate
these as important topics for improvement. 

In a forthcoming paper we will study the other four bursts from \source{}.
Anticipating improvements in \emph{NICER}'s gain calibration and response
model, we will search for discrete spectral features from the neutron
star surface and the wind. Moreover, we will investigate constraints
on the neutron star's mass and radius that can be derived from the
PRE phase. Further \emph{NICER} observations of \source{} may catch
a burst at lower persistent flux, where the mass accretion inflow
is smaller than the wind outflow, leading to different expansion behavior
and the possibility of detecting redshifted absorption features \citep{Zand2012}.

\acknowledgements{This work was supported by NASA through the \emph{NICER} mission
and the Astrophysics Explorers Program. This research has made use
of data and software provided by the High Energy Astrophysics Science
Archive Research Center (HEASARC), a service of the Astrophysics Science
Division at NASA/GSFC and the High Energy Astrophysics Division of
the Smithsonian Astrophysical Observatory. This work benefited from
events supported by the National Science Foundation under grant No.
PHY-1430152 (JINA Center for the Evolution of the Elements). S.G.
acknowledges CNES.}

\bibliographystyle{apj}
\bibliography{1820nicer}

\end{document}